\documentclass[12pt]{iopart}

\usepackage{enumitem}
\usepackage{natbib}
\usepackage{amssymb}
\usepackage{graphicx}
\usepackage{color}
\usepackage{amsthm}
\usepackage{slashed}
\usepackage{soul}
\usepackage{amsbsy}

\begin{document}

\title[Weyl-Invariant Gravity and the Nature of Dark Matter]
{Weyl-Invariant Gravity and the Nature of Dark Matter}

\author{Meir Shimon}

\address{School of Physics and Astronomy, Tel Aviv University, Tel Aviv 69978, Israel}
\ead{meirs@tauex.tau.ac.il}
\vspace{10pt}

\begin{abstract}
The apparent missing mass in galaxies and galaxy clusters, 
commonly viewed as evidence for dark matter, could possibly originate from 
gradients in the gravitational coupling parameter, $G$, and active gravitational 
mass, $M_{act}$, rather than hypothetical beyond-the-standard-model particles. 
We argue that in (the weak field limit of) a Weyl-invariant extension of General 
Relativity, one can simply affect the change 
$\Phi_{b}(x)\rightarrow\Phi_{b}(x) + \Phi_{DM}(x)$, where $\Phi_{b}$ is 
the baryon-sourced potential and $\Phi_{DM}$ is the `excess' potential. This is 
compensated by gradients of $GM_{act}$ and a fractional increase of 
$O(-4\Phi_{DM}(x))$ in the baryon density, well below current detection thresholds 
on all relevant scales. 
\end{abstract}

%\maketitle

\section{Introduction}

Two key components of the standard model (SM) of cosmology, $\Lambda$CDM, 
are dark energy (DE) and dark matter (DM), where evidence for DM 
on galactic and cosmological scales abounds.
It is commonly thought that observations favor a type of `cold' DM (CDM) on 
all relevant scales. 
CDM constitutes $\sim 80\%$ of the non-relativistic (NR) matter in 
the universe, and together with DE makes up the `dark sector' of cosmology, 
accounting for $\sim 95\%$ of the energy budget of the universe at present. 

DM has reincarnated several times over the nearly century old 
history of the modern era of cosmology [1-3]. 
The first indication for missing mass, i.e. DM, in galaxy clusters 
was provided by the velocity dispersion of a handful of galaxies in 
the Coma cluster [4, 5]. Five decades later it has been concluded 
based on observations of rotation curves in galaxies that DM halos 
dominate the mass budget on galactic scales [6].  At around the same 
time a $\sim$1 Kev mass neutrino was 
proposed as the dominant component of the NR energy budget on 
cosmological scales [7]. 

Obvious DM candidates are massive compact halo objects (MACHOs) such as 
primordial black holes [8-11], but this possibility seems to be less 
popular at present as compared to weakly interacting massive particles (WIMPs), 
e.g. [12-19]. 
The WIMP paradigm is held by many as the most attractive microphysical 
explanation for DM on galactic and extragalactic scales. 
Yet, no definite detection had been made in any 
of the few dozen terrestrial DM particle detection experiments.

Cold, warm, and hot DM candidates 
are distinguished by the impact they have on structure formation hierarchy and are thus 
constrained by probes of the growth of structure. The simplest hot DM 
candidates are already ruled out on galactic scales. Warm DM 
candidates are thought to be either gravitinos or sterile neutrinos [20-22] 
that were non-thermally produced unlike the leading CDM particles 
which would account for WIMPs that are thermally produced.
While CDM is the leading candidate on Hubble down to galactic scales, certain 
issues with CDM on dwarf galaxy scales seem to be more optimally addressed by 
fuzzy DM, e.g. [23]. Mimetic DM 
is yet another DM model that provides a {\it geometrical} (rather than material) 
interpretation for DM on cosmological scales, e.g. [24]. 
An alternative geometrical interpretation 
is provided by Horava gravity [25].

The allowed parameter space of DM particles ranges from ultralight to 
superheavy, that can be either bosons or fermions, neutral or charged, stable or unstable, 
interacting with DE or not, etc [26-29]. Inspired by supersymmetry, the list of candidate DM particles is large, e.g. [30]. These include: the neutralino, axion, 
sterile neutralino, gravitinos, axino, Q-balls, etc  [31-33]. 
The number of model parameters in the minimal supersymmetric standard model 
(MSSM) exceeds 100, and under a few simplifying assumptions this number can be 
considerably reduced to $\sim 20$. Despite this ample freedom, there is 
yet no single candidate that fully explains the entire spectrum of DM phenomena, 
especially on the smallest cosmological scales, e.g. [34, 35]. 
In spite of extensive theoretical effort there is no viable unified DM model.

With dozens of direct [36-39] and indirect [40-46] detection experiments, 
as well as with the large hadron collider (LHC), a vigorous 
hunt for WIMPs, non-WIMP or Kaluza-Klein 
states [47, 48] is underway. Present and future indirect efforts aim at detecting 
either $\gamma$-ray emission from DM annihilation at the Galactic center 
and dwarf galaxies or via high energy neutrinos from the Sun [49-54].

Perhaps the first compelling alternative approach to  
explain galactic rotation curves and velocity dispersion in galaxy clusters was 
modified Newtonian dynamics (MOND) [55-58]. As a phenomenological model,
MOND lacks a relativistic description thus preventing its application on cosmological 
scales. Relativistic theories of modified gravity (MOG) and mimetic gravity, 
also aim at explaining galactic rotation curves with 
no recourse to particle DM [59, 60]. 
These typically include both scalar, vector 
and tensor degrees of freedom [61, 62].

An alternative Weyl-invariant (WI) relativistic 
theory -- fourth-order Weyl gravity -- was later proposed to account for the anomalous rotation curves in galaxies with no recourse to DM [63-66]. 
The proposed remedy is 
based on an exact spherically symmetric static solution of the field equations. 
A similar solution based on Weyl invariant scalar tensor (WIST) gravity was found in [67]. However, it has not been demonstrated that these 
alternative approaches provide compelling explanations in other systems, e.g. galaxy clusters, and in more realistic (dynamical and non--symmetric) cases such as merging bullet-like galaxy clusters.
Other attempts to resolve the DM enigma include, in particular, scalar-tensor (ST)  
theories that provide {\it specific} mechanisms with falsifiable predictions.

The discovery of a few bullet-like merging galaxy clusters is seen 
by many as a serious challenge to MOND and other alternatives to the DM paradigm [68] . In these systems the 
center of gravity (probed by lensing) and the luminous baryonic matter (observed through its X-ray emission) are clearly separated and this is conventionally interpreted as 
evidence that the bulk mass of the merging clusters is made of dissipasionless matter.
However, the existence of DM in these systems is deduced 
from lensing measurements (of background quasars). 
The latter depends on transversal gradients of the gravitational potential 
$\nabla_{\perp}\Phi$; the equivalent DM density profile $\rho_{DM}(x)$, which is not directly measured, can be deduced from maps 
of $\nabla_{\perp}\Phi$ (integrated along the line of sight). 

Crucial to this procedure in particular, and to the CDM paradigm in general, is the 
assumption that the gravitational coupling $G$ is a universal constant. 
This has been established to a high level of precision essentially only with regard 
to its {\it temporal} evolution, e.g. [69, 70]. 
However, more relevant to the current work is a possible space-dependence of 
the Planck mass, $m_{P}\propto G^{-1/2}$, 
and more specifically scale-dependence, if an underlying translation 
invariance is assumed. In comparison, the observed scale-dependence 
of DM phenomena is implicitly set in WIMP or fuzzy DM 
models by, e.g. the particle mass or de Broglie wavelength, respectively.    
We note in passing that even if the conservative interpretation of 
bullet-like clusters is accepted at face value, the existence of 
these systems only implies that DM cannot be fully accounted for 
by baryons, by no means does this prove that DM is {\it particulate}.

\section{WIST theory}

Our universe is highly symmetric; the electroweak and strong interactions 
are described by a $U(1)\otimes SU(2)\otimes SU(3)$ gauge symmetry, and the underlying 
symmetry of general relativity (GR) is general coordinate invariance. 
A hitherto hidden symmetry of GR 
could be WI. The ultimate test of the validity of symmetries 
in physical theories is experiment, as well as naturalness, falsifiability, etc.

While the assumptions that the Higgs mass scale $v\approx 246$ GeV and 
chiral symmetry breaking mass scale $\Lambda_{QCD}\approx 220$ MeV are 
universally fixed seem to work well for the SM of particle physics, 
a similar assumption for $G$ may fail already on 
galactic scales (unless DM is invoked or gravity is modified), although 
it is consistent with solar system observations. 
If $G$ is determined by a scalar field then 
the convention that the former is universal 
obviously implies constancy of the latter. 
In the following, we relax this assumption and allow variation of $GM_{act}$, 
where $M_{act}$ is the active gravitational mass, in space and time [71]. 
The equivalence of passive and inertial masses is 
essentially the `equivalence principle' which is not invalidated 
by allowing for variations of $GM_{act}$ [72].

The WIST theory considered here is 
obtained by replacing the spacetime metric $g_{\mu\nu}$ with $\phi\phi^{*}g_{\mu\nu}$ 
everywhere in the Einstein-Hilbert (EH) action 
$\mathcal{I}_{EH}=\int[R/(16\pi G)+\mathcal{L}_{m}]\sqrt{-g}d^{4}x$ (in units 
where $G\equiv\frac{3}{8\pi}$), 
where $\phi$ \& $\phi^{*}$ are a scalar field and its complex conjugate, 
$\mathcal{L}_{m}$ is the lagrangian density of matter and $g$ is the 
metric determinant. The resulting ST theory is
\begin{eqnarray}
\mathcal{I}_{ST}&=&\int\left(\frac{1}{6}|\phi|^{2}R
+\phi_{\mu}\phi^{*\mu}+\mathcal{L}_{m}(|\phi|,\{\psi\})\right)\sqrt{-g}d^{4}x\nonumber\\
&=&\int\left(\frac{1}{6}|\phi|^{2}R
-\phi^{*}\Box\phi+\mathcal{L}_{m}(|\phi|,\{\psi\})\right)\sqrt{-g}d^{4}x,
\end{eqnarray}
where $\phi_{\mu}\equiv\phi_{,\mu}$, $\mathcal{L}_{m}$ is now 
allowed to explicitly depend on $|\phi|$ but not on its derivatives, 
and the second equality follows from integration by parts 
of the canonical kinetic term associated with the scalar field.
All other fields are collectively denoted by $\{\psi\}$.
The kinetic term associated with the scalar field 
$\mathcal{L}_{\phi}\equiv -\phi^{*}\Box\phi$ can be 
considered as a new source of the gravitational force. This 
term is completely ignored in GR as $|\phi|$ is a constant 
$\sqrt{\frac{3}{8\pi G}}$ in this case.

The model has a global U(1) symmetry but we will not need the phase of the 
scalar field here. The latter plays an important role in the very early 
universe in a cosmological model based on Eq. (1) [73]. 
By construction, the action is invariant under Weyl 
transformations, i.e. $g_{\mu\nu}\rightarrow \Omega^{2}g_{\mu\nu}$, 
$\mathcal{L}_{m}\rightarrow\Omega^{-4}\mathcal{L}_{m}$, $\phi\rightarrow\phi/\Omega$,  
with $\Omega(x)$ a continuous but otherwise arbitrary spacetime-dependent function.
The theory described by Eq. (1) was first considered in [74]  
for the case of real $\phi$ and $\mathcal{L}_{m}$ independent of $\phi$.
This additional freedom to locally rescale fields is essential to our construction; 
it implies that $\Omega(x)$ can be so chosen so as to modify any 
single metric perturbation potential to obtain any desired 
value at any point in spacetime, but at the cost 
of corresponding modifications to $\phi$ (i.e. $G$), and 
$\mathcal{L}_{m}$ (that 
determine the energy density of matter and its pressure). 
This is exactly the case we consider below; neglecting any 
vector and tensor perturbation modes, and assuming vanishing stress, 
metric perturbations of scalar type are described by a single function $\Phi(x)$.

Variation of Eq. (1) with respect to the metric and $\phi^{*}$ results 
in the following field equations, respectively, e.g. [75] 
\begin{eqnarray}
\frac{|\phi|^{2}}{3}G_{\mu\nu}=T_{M,\mu\nu}+\Theta_{\mu\nu}\\
\frac{1}{6}\phi R-\Box\phi+\frac{\partial\mathcal{L}_{m}}{\partial\phi^{*}}=0,
\end{eqnarray}
where
\begin{eqnarray}
3\Theta_{\mu\nu}\equiv\phi^{*}_{\mu;\nu}\phi-2\phi^{*}_{\mu}\phi_{\nu}
-g_{\mu\nu}(\phi^{*}\Box\phi-\frac{1}{2}\phi_{\alpha}^{*}\phi^{\alpha})+c.c. 
\end{eqnarray}
Eq. (2) is a generalization of Einstein equations, 
and $\Theta_{\mu\nu}$ is an effective contribution to 
the energy-momentum tensor essentially due to gradients 
of $\phi$ (that determines $G$ and $M_{act}$).
From Eq. (4) it then follows that the trace 
of $\Theta_{\mu\nu}$ is $\Theta=\mathcal{L}_{\phi}$.

In the weak field limit Eq. (2) results in a 
modified Poisson equation.
Multiplying Eq. (3) by $\phi^{*}$ and adding the result to 
its complex conjugate and to the trace of Eq. (2) results in the constraint
\begin{eqnarray}
\phi^{*}\frac{\partial\mathcal{L}_{m}}{\partial\phi^{*}}
+\phi\frac{\partial\mathcal{L}_{m}}{\partial\phi}=T_{m}, 
\end{eqnarray}
i.e. only pure radiation $T_{rad}=0$ is consistent with WI gravity unless 
$\mathcal{L}_{m}$ (which in the case of perfect fluid equals $T_{m}$) 
explicitly depends on $|\phi|$. This fact explains 
the requirement in Eq. (1) that in general the matter lagrangian 
does depend on $|\phi|$.
Recalling that $\mathcal{L}_{m}$ is a potential in $\phi$, 
it then follows that 
for NR matter, $\rho_{NR}\propto\rho$, i.e. $M_{act}\propto\rho$, where 
$\rho$ is the modulus of $\phi$. 
For a general equation of state (EOS) $w\equiv P_{m}/\rho_{m}$, 
where $P_{m}$ 
is the pressure associated with matter of energy density $\rho_{m}$, 
Eq. (5) is satisfied by  
$\mathcal{L}_{m}\propto\rho^{1-3w}=(\phi\phi^{*})^{\frac{1-3w}{2}}$.
We mention in passing that no violation of the equivalence principle ensues insofar inertial 
and passive gravitational masses are fixed, which we indeed assume here. 
The notion of passive gravitational masses is vacuous in the present theory and 
we only use the 
Newtonian parlance for clarity. Irrespective of that, the equivalence principle 
has never been directly tested beyond the solar system anyway, and obviously 
not directly with DM particles.

The relevant 
combination for NR gravitating source is $GM_{act}$ which is 
$\propto\rho^{-1}$. 
While $GM_{act}$ is spacetime-dependent, 
we assume that the SM of particle physics 
is left unchanged (i.e. WI is not a symmetry thereof), 
and inertial masses are universal constants.
 
\section{DM in gravitationally-bound structures}
A key to the following considerations is the fact that 
`DM' is only probed via the force its exerts, 
rather than $\rho_{DM}$. The latter, we 
recall, is absent from $\mathcal{L}_{m}$ in Eq. (1).
In contrast, density distribution of baryonic matter, 
$\rho_{b}$, is directly observed through electromagnetic 
emission, absorption and scattering processes. 
Throughout, we assume that DM is not real matter and that 
its apparent attributes arise from gradients of $GM_{act}$.

Although we focus on WIST, the following 
arguments are equally well valid in any WI theory of gravitation. 
However, in general other theories do not have a Poisson-like equation as 
their weak field limit, and thus $G$ is only an effective notion, e.g. [76]. 
Although there is no scalar field in the gravitational sector of forth-order 
Weyl gravity such a scalar (or an effective scalar) field is expected to 
appear in $\mathcal{L}_{m}$. The latter must scale $\propto\Omega^{-4}$ 
in any WI theory, and since there are no mass scales in the theory, 
effective masses, e.g. the Planck mass or active gravitational mass 
(the mass that sources the gravitational field in $\mathcal{L}_{m}$ that 
need not be identical with the passive gravitational mass or the inertial mass), 
scale $\propto\Omega^{-1}$. 

Neglecting DM effects, 
assuming negligible stress and vanishing vector and tensor 
modes, the appropriate spacetime for 
galaxies and galaxy clusters is described 
(in the weak field limit) by the line element 
\begin{eqnarray}
ds^{2}=-(1-2\Phi_{b})dt^{2}-dtdx^{i}v_{i}+(1+2\Phi_{b})dx_{i}dx^{i}, 
\end{eqnarray}
where Latin indices (denoting 
spatial coordinates) are summed over, $v_{i}\ll 1$ is a NR velocity and 
$\Phi_{b}\ll 1$ is the linear gravitational potential that satisfies the 
relativistic generalization of Poisson equation with a source term $\rho_{b}$.  
Here, $v_{i}$ is already a linear perturbation over Minkowski spacetime. 
Other than the condition $\Phi_{b}\ll 1$ we make no 
specific assumptions about the spacetime-dependence (symmetry) of $\Phi_{b}$ 
and the following results are therefore general.  
Linear analysis is justified by the fact that the 
rms internal velocities of substructure within galaxies and 
galaxy clusters are much lower than the speed of light, i.e. 
$\Phi_{b}\ll 1$ \& $\Phi_{DM}\ll 1$.

Applying a Weyl transformation with $\Omega=1+\Phi_{DM}(x)$ and recalling that 
$g_{\mu\nu}\rightarrow\Omega^{2}g_{\mu\nu}$ and $\phi\rightarrow\phi/\Omega$,
then at {\it first order}, the gravitational potential 
in Eq. (6) is shifted, $\Phi_{b}\rightarrow\Phi_{b}+\Phi_{DM}$, at 
the ``cost'' of $GM_{act}\propto 1+\Phi_{DM}$ and a corresponding change 
in $\rho_{b}$ that we discuss below. 

We assume that $\Phi_{DM}\ll 1$ on the relevant range of scales 
but is {\it a priori} an {\it arbitrary} function of spacetime. 
It accounts for the excess 
gravitational potential required to explain galactic 
scale observations. Translation invariance further implies 
that $\Omega$ is a function of scale rather than location. 
Specifically, in the spherically symmetric static case, we could 
select the Navarro-Frenk-White (NFW) [77] 
gravitational potential profile, $\Phi_{NFW}(r)\propto r^{-1}\ln(1+r/r_{s})$ 
(where $r_{s}$ is an object-specific scale parameter), or any other profile, 
e.g. [78], as our $\Phi_{DM}(r)$.
The underlying assumptions about the microphysics that 
lead to the specific r-dependence 
of $\Phi_{NFW}(r)$ (or any other $\Phi_{DM}(r)$ profile 
for that matter that fits observational data) 
are irrelevant to our discussion.  
It only matters that the combination 
$\Phi=\Phi_{b}(x)+\Phi_{DM}(x)$ provides a good fit to the data.
The excess gravitational potential $\Phi_{DM}$ is supported 
by gradients of $\phi$, i.e. of $GM_{act}$, 
via the $\Theta_{\mu\nu}$ addition 
to the baryonic energy-momentum tensor in Eq. (2). 
In this sense, ``$T_{DM,\mu\nu}$''$=\Theta_{\mu\nu}$. 
Its nonvanishing components in spherical coordinates, 
linear in $\Phi_{DM}\ll1$, 
are $\Theta_{\eta}^{\eta}=\frac{\phi^{2}}{3}(\Phi_{rr}+\frac{2}{r}\Phi_{r})$, 
$\Theta_{r}^{r}=\frac{2\phi^{2}}{3r}\Phi_{r}$ 
and $\Theta_{\theta}^{\theta}=\Theta_{\varphi}^{\varphi}
=\frac{\phi^{2}}{3}(\Phi_{rr}+\frac{1}{r}\Phi_{r})$, 
where the `DM' subscript was dropped and it is assumed that 
$\Phi=\Phi(r)$. 
Interestingly, the trace $\rho_{DM}-3P_{DM}=-\Theta=3\rho_{DM}$, 
which corresponds to an effective (averaged over spatial directions) 
EOS $-2/3$. Although $w_{DM}=-2/3$ 
is negative and is very nearly 
the EOS of the expanding Universe at the present, 
we believe that this is merely a coincidence. Similarly irrelevant 
to the present work is the fact that it is the same EOS of 
a `domain wall' solution of the Einstein equations.
The pressure in the r-direction differs in general 
from the pressure in the other two orthogonal directions. 
Therefore, the {\it effective} energy-momentum tensor 
associated with DM does not correspond to a perfect fluid, 
and in any case does not describe pressureless matter as does CDM.
This is not a problem since neither DM pressure nor DM density 
[and consequently not its EOS $w_{DM}$] 
are directly observed. Again, in practice the CDM paradigm 
with $w_{DM}\approx 0$ is only a means to obtain an adequate $\Phi_{DM}(r)$ fit 
on these scales assuming the latter satisfies the Poisson, Euler and continuity equations 
{\it provided that G is a universal constant}. These equations are essentially obtained 
from Eq. (2) 
and its derivatives in the more general case (varying $G$ and $M_{act}$) considered here.
As emphasized above, from the WI perspective any {\it guess}, 
or e.g. artificial intelligence informed reconstruction,
of $\Phi_{DM}(r)$ is equally legitimate insofar as it provides a 
reasonably good fit to the data combined with 
$\Phi_{b}$, much like $|\phi|^{2}=3/(8\pi G)=constant$ across the entire universe 
and over Hubble time is an educated guess based on observations over the relatively 
small solar system scales. 
However, a significant difference is that in the latter convention DM is required 
(to make up for the convenient $|\phi|^{2}=3/(8\pi G)=constant$ choice) which is 
widely believed to consist of beyond-the-SM particles.

Aside from accounting for the apparent effects of DM, 
the only (potentially observational) consequence of the $\Omega=1+\Phi_{DM}(x)$ 
choice in these systems is that $\rho_{b}$ transforms as 
$\rho_{b}\rightarrow\rho_{b}/(1+\Phi_{DM})^{4}\approx\rho_{b}(1-4\Phi_{DM})$.
Since typically $-\Phi_{DM}=O(10^{-4})$ in galaxies or $O(10^{-5})$ in 
galaxy clusters, then $O(4\Phi_{DM})$
amounts to a tenth of a percent increase in $\rho_{b}$ at most, 
too weak to be measured given the typically complicated morphology 
and lumpiness of baryon distribution, temperature uncertainty, 
current observational precision, etc, on these scales. 
Therefore, the transformation $\Omega=1+\Phi_{DM}(x)$ can significantly modify the 
gravitational potential, e.g. outside galaxy cluster and galaxy cores, while making only a 
subtle change in 
the observable luminous matter density, thereby making the case for 
a `non-particle DM'. This procedure could clearly be applied 
to merging bullet-like clusters as well, where in this case $\Phi_{DM}$ 
is expected to depend on both space and time, with no impact on our general 
conclusion that DM may well be a non-WIMP, and non-MACHO-related 
phenomenon on galactic and galaxy cluster scales.  

The approach explored here to remove the need for DM 
on galactic and super-galactic scales could be 
similarly employed on cosmological scales since typically $\Phi=O(10^{-5})$ 
on these scales. However, a NR DM component seems to still be required at 
the background level by a wealth of cosmological probes. In the concordance 
cosmological model the NR component accounts for $\sim 30\%$ of the total  
energy density at present, while baryons account for only $\sim 5\%$; 
the remaining $\sim 25\%$ is believed to be in CDM. 

Similar to the replacement of $g_{\mu\nu}$ with $\phi\phi^{*}g_{\mu\nu}$ that 
leads to Eq. (1), replacing the metric with 
$g^{\alpha\beta}\varphi_{\alpha}\varphi_{\beta}g_{\mu\nu}$ 
(along with the constraint 
$g^{\alpha\beta}\varphi_{\alpha}\varphi_{\beta}=1$, 
where $\varphi$ is a real scalar field) 
in the EH action could be made, that 
leads to ``mimetic gravity''. The latter theory was 
proposed as a solution to the DM problem on the largest cosmological scales 
assuming the homogeneous and isotropic Friedmann-Robertson background metric 
[79, 80]. According to this picture the additional energy provided 
by $\varphi_{\mu}$ mimics CDM with an EOS 
$w_{mim}=0$. This does not contradict our proposed solution 
in gravitationally bound structures for which the effective matter 
density and pressure do not correspond with NR matter since 
the ``missing mass'' problems in these two entirely different 
physical environments are of completely different nature; $GM_{act}$ is 
indeed constant on cosmological scales as in the SM of cosmology, but 
spatially varies on galactic scales.

\section{Summary}

The perspective advocated in the present work is that 
WI may well be a symmetry of gravitation, and that while GR represents a convenient 
specific choice of units (namely constant $G$ and $M_{act}$), it may have 
misled us to think that some exotic form of DM is required to account for observations (either 
in the form of MACHOs or WIMPs). A more prosaic solution to the DM problem might 
be that $GM_{act}$ is actually scale-dependent.

In the context of the proposed ST theory 
gradients of the scalar field (essentially of $GM_{act}$)
may provide the required additional energy and pressure 
for warping space to reproduce the observed gravitational pull caused by hypothesized 
DM without recourse to DM particles. Specifically, we show that 
DM manifestations could be fully accounted for through the innocuous change 
$\Phi_{b}\rightarrow \Phi_{b}+\Phi_{DM}$ at the `cost' of $O(\Phi_{DM})$ 
fractional variations in $GM_{act}$ and $\rho_{b}$ 
in the range of $O(10^{-3})-O(10^{-4})$ 
on the relevant galactic and super-galactic 
scales, levels that are too weak 
to be directly observed.

As the conformal factor $\Omega(x)=1+\Phi_{DM}(x)$ 
[and consequently $\Phi_{DM}(x)$] is arbitrary, it could be 
in principle selected to better-fit observations (on all relevant 
scales, including dwarf galaxy scales) than would standard particle-induced 
CDM gravitational potential profiles do. 
We note that at the cosmological background level 
GR endowed with WI is equivalent to mimetic gravity, and thus contains 
DM-like energy density and energy density perturbations with no recourse to DM particles.    

It may seem that since the conformal 
factor $\Omega(x)=1+\Phi_{DM}(x)$, essentially $GM_{act}(x)$, can always 
be so chosen to fit the data then the theory is not falsifiable. 
However, by the same rationale the choice $|\phi|^{2}=3/(8\pi G)=constant$ 
employed in the SM has already been falsified by observations on galactic scales, 
and it is only saved by invoking an unobservable $\rho_{DM}(x)$ component. 
It is somewhat ironic that if existing DM candidates are not found, 
the CDM paradigm may still hold up with the only consequence that 
the vast WIMP parameter space is narrowed down, while in contrast, 
if WIMPs are found with properties that are sufficient 
for adequately explaining observations, our proposed solution 
(as is any other alternative solution) to the 
DM problem is essentially falsified.

We conclude that if gravitation is indeed locally scale-invariant, 
a symmetry hidden by universally setting $|\phi|^{2}=3/(8\pi G)=constant$ 
in GR (largely motivated by convenience, at least on galactic and cosmological scales),  
then the currently favorite identification of DM with 
WIMPs may just be an artifact of arbitrarily adopting a constant system of units.

\section*{Acknowledgments}
The author is indebted to Yoel Rephaeli for numerous constructive 
and useful discussions, and for his continued support and encouragement.
Anonymous referees of this work are also acknowledged 
for their useful comments.
This research has been supported by a grant from 
the Joan and Irwin Jacobs donor-advised fund at the JCF (San Diego, CA).

\section*{References}
\begin{enumerate}[label={[\arabic*]}]
\item Bertone G and Hooper D 2018 {\it Rev. Mod. Phys.} {\bf 90} 045002
\item Trimble V 1987 {\it Annu. Rev. Astron. Astrophys.} {\bf 25} 425
\item Peebles P.~J.~E 2017 {\it Nat. Astron.} {\bf 1} 0057
\item Zwicky F 1933 {\it Helv. Phys. Acta} {\bf 6} 110
\item Andernach H and Zwicky F 2017 {\it arXiv:1711.01693}
\item Rubin V.~C 1983 {\it Science} {\bf 220} 1339
\item Peebles P.~J.~E 1982 {\it Astrophys. J. Lett.} {\bf 263} L1
\item Chapline G.~F 1975 {\it Nature} {\bf 253} 251
\item Carr B, K{\"u}hnel F and Sandstad M 2016 {\it Phys. Rev.} D {\bf 94}, 083504
\item Carr B and K{\"u}hnel F 2020 {\it Annu. Rev. Nucl. Part. Sci.} {\bf 70} 
\item Frampton P.~H 2016 {\it Mod. Phys. Lett. A} {\bf 31} 1650093
\item Bergstr{\"o}m L 2000 {\it Rep. Prog. Phys.} {\bf 63} 793
\item Griest K and Kamionkowski M 2000 {\it Phys. Rep.} {\bf 333} 167
\item Mu{\~n}oz C 2004 {\it Int. J. Mod. Phys.} A {\bf 19} 3093
\item Bergstr{\"o}m L 2009 {\it New J. Phys.} {\bf 11} 105006
\item Feng J.~L 2010 {\it Annu. Rev. Astron. Astrophys.} {\bf 48} 495 
\item Steigman G, Dasgupta B and Beacom J.~F 2012, {\it Phys. Rev.} D {\bf 86} 023506
\item Roszkowski L, Sessolo E.~M and Trojanowski S 2018 
{\it Rep. Prog. Phys.} {\bf 81} 066201
\item Jungman G, Kamionkowski M and Griest K 1996, 
{\it Phys. Rep.} {\bf 267} 195
\item Adhikari R Agostini M Ky N.~A. et al. 2017 {\it J. Cosmol. Astropart. Phys.} 2017 025
\item Boyarsky A, Drewes M, Lasserre T, et al. 2019, 
{\it Prog. Part. Nucl. Phys.} {\bf 104} 1
\item Kusenko A 2009 {\it Phys. Rep.} {\bf 481} 1
\item Hu W, Barkana R and Gruzinov A 2000 {\it Phys. Rev. Lett.} {\bf 85} 1158
\item Chamseddine A.~H and Mukhanov V 2013 {\it J. High Energy Phys.} JHEP11(2013)135
\item Mukohyama S 2009 {\it Phys. Rev.} D {\bf 80} 064005
\item Chung D.~J.~H, Kolb E.~W and Riotto A 1999 {\it Phys. Rev.} D {\bf 59} 023501
\item Ferreira E.~G.~M 2020 {\it arXiv:2005.03254}
\item de R{\'u}jula A, Glashow S.~L and Sarid U 1990 {Nucl. Phys.} B {\bf 333} 173
\item Dimopoulos S, Eichler D, Esmailzadeh R, et al. 1990, 
{\it Phys. Rev.} D {\bf 41}, 2388
\item Bertone G 2010 {\it Particle Dark Matter : Observations, Models and Searches, Edited by Gianfranco Bertone. Published by Cambridge University Press, Cambridge, UK ; New York :  2010. ISBN: 9780521763684}
\item Marsh D.~J.~E 2016 {Phys. Rep.} {\bf 643} 1
\item Covi L, Kim J.~E and Roszkowski L 1999 {\it Phys. Rev. Lett.} {\bf 82} 4180
\item Kusenko A and Shaposhnikov M 1998 {\it Phys. Lett.} B {\bf 418} 46
\item Bullock J.~S and Boylan-Kolchin M 2017 {\it Annu. Rev. Astron. Astrophys.} {\bf 55} 343
\item Milgrom M 2020 {\it Studies in the History and Philosophy of Modern Physics} {\bf 71} 170
\item Aprile E, Aalbers J, Agostini F, et al. 2017 {\it Phys. Rev. Lett.} {\bf 119} 181301
\item Ahmed Z, Akerib D.~S, Arrenberg S, et al. 2009 {\it Phys. Rev. Lett.} {\bf 102} 011301
\item Aprile E, Aalbers J, Agostini F, et al. 2016 {\it J. Cosmol. Astropart. Phys.} {\bf 2016} 027
\item Aalbers, J., Agostini, F., Alfonsi, M., et al.\ 2016 {\it J. Cosmol. Astropart. Phys.} {\bf 2016} 017
\item Adriani O, Barbarino G.~C, Bazilevskaya G.~A, et al.\ 2009 {\it Phys. Rev. Lett.} {\bf 102} 051101
\item Geringer-Sameth A and Koushiappas S.~M\ 2011, 
{\it Phys. Rev. Lett.} {\bf 107} 241303
\item Bergstr{\"o}m L, Bringmann T and Edsj{\"o} J\ 2008 {\it Phys. Rev.} D {\bf 78} 103520
\item Bergstr{\"o}m L, Ullio P and Buckley J.~H\ 1998 {\it Astropart. Phys.} {\bf 9} 137
\item Ackermann M, Ajello M, Allafort A, et al.\ 2010 {\it J. Cosmol. Astropart. Phys.}
{\bf 2010} 025
\item Cesarini A, Fucito F, Lionetto A, et al.\ 2004
{\it Astropart. Phys.} {\bf 21} 267
\item Abazajian K.~N and Kaplinghat M\ 2012 {\it Phys. Rev.} D {\bf 86} 083511 
\item Cheng H.-C, Feng J.~L and Matchev K.~T\ 2002 {\it Phys. Rev. Lett.} {\bf 89} 211301
\item Hooper D and Kribs G.~D\ 2004 {\it Phys. Rev.} D 70
\item Bertone G, Hooper D and Silk J\ 2005, 
{\it Phys. Rep.} {\bf 405} 279
\item Bergstr{\"o}m L\ 2012 {\it Ann. Phys.} {\bf 524} 479
\item Arcadi G, Dutra M, Ghosh P, et al.\ 2018, {\it Eur. Phys. J.} C {\bf 78} 203
\item Ackermann M, Albert A, Anderson B, et al.\ 2015 {\it Phys. Rev. Lett.} {\bf 115} 231301
\item Ullio P, Bergstr{\"o}m L, Edsj{\"o} J, et al.\ 2002, {\it Phys. Rev.} D {\bf 66} 123502
\item Abdo A.~A, Ackermann M, Ajello M, et al.\ 2010 {\it Astrophys. J.} {\bf 712} 147
\item Milgrom M\ 1983 {\it Astrophys. J.} {\bf 270} 365
\item Milgrom M\ 1983 {\it Astrophys. J.} {\bf 270} 371
\item Milgrom M\ 2010 {\it Invisible Universe} {\bf 1241} 139
\item Bekenstein J and Milgrom M\ 1984 {\it Astrophys. J.} {\bf 286} 7
\item Moffat J.~W and Rahvar S\ 2013 {\it Mon. Not. R. Astron. Soc.} {\bf 436} 1439
\item Myrzakulov R, Sebastiani L, Vagnozzi S, et al.\ 2016 {\it Class. Quantum Gravity} {\bf 33} 125005
\item Bekenstein J.~D\ 2004 {\it Phys. Rev.} D {\bf 70} 083509
\item Moffat J.~W\ 2006 {\it J. Cosmol. Astropart. Phys.} {\bf 2006} 004
\item Mannheim P.~D and Kazanas D\ 1989 {\it Astrophys. J.} {\bf 342} 635
\item Mannheim P.~D and O'Brien J.~G\ 2011 {\it Phys. Rev. Lett} {\bf 106} 121101
\item Mannheim P.~D and O'Brien J.~G\ 2012 {\it Phys. Rev.} D {\bf 70} {\bf 85} 124020 
\item O'Brien J.~G and Mannheim P.~D\ 2012 {\it Mon. Not. R. Astron. Soc.} {\bf 421} 1273
\item Li Q and Modesto L\ 2020 {\it Gravit. Cosmol.} {\bf 26} 99
\item Clowe D, Brada{\v{c}} M, Gonzalez A.~H, et al.\ 2006 {\it Astrophys. J. Lett} {\bf 648} L109
\item Bertotti B, Iess L and Tortora P\ 2003 {\it Nature} {\bf 425} 374
\item Zhu W.~W, Stairs I.~H, Demorest, P.~B, et al.\ 2015 {\it Astrophys. J.} {\bf 809} 41
\item Bondi H.\ 1957 {\it Reviews of Modern Physics} {\bf 29} 423
\item Roll, P.~G., Krotkov, R., \& Dicke, R.~H.\ 1964 {\it Annals of Physics} 
{\bf 26} 442
\item Shimon M\ 2021, in preparation
\item Deser S\ 1970 {\it Ann. Phys.} {\bf 59} 248
\item Hwang, J.-C. \& Noh, H.\ 2002 {\it Phys. Rev.} D {\bf 65} 023512
\item Mannheim P.~D and Kazanas D\ 1994 {\it Gen. Relativ. Gravit.} {\bf 26} 337
\item Navarro J.~F Frenk C.~S and White S.~D.~M\ 1997 {\it Astrophys. J.} {\bf 490} 493
\item Merritt D, Graham A.~W, Moore B, et al.\ 2006 {\it Astron. J.} {\bf 132} 2685
\item Chamseddine A.~H and Mukhanov V\ 2013 {\it J. High Energy Phys.} {\bf 2013} 135
\item Matsumoto J, Odintsov S.~D and Sushkov S.~V\ 2015 {\it Phys. Rev.} D {\bf 91} 064062
\end{enumerate}

\end{document}